\documentclass[aps,prd,amsmath,amssymb,reprint,superscriptaddress]{revtex4-1}
\usepackage{graphicx}
\usepackage{amsmath}
\usepackage{dcolumn}
\usepackage{bm}
\usepackage{bbold}
\usepackage{color}
\usepackage{amsfonts}
\usepackage{amssymb}
\usepackage{mathrsfs}
\usepackage{braket}

\usepackage{caption}
\usepackage{subcaption}

\begin{document}  

\title{Gravitational Wave Detection with High Frequency Phonon Trapping Acoustic Cavities}

\author{Maxim Goryachev}
\affiliation{ARC Centre of Excellence for Engineered Quantum Systems, School of Physics, University of Western Australia, 35 Stirling Highway, Crawley WA 6009, Australia}

\author{Michael E. Tobar}
\email{michael.tobar@uwa.edu.au}
\affiliation{ARC Centre of Excellence for Engineered Quantum Systems, School of Physics, University of Western Australia, 35 Stirling Highway, Crawley WA 6009, Australia}

\date{\today}


\begin{abstract}

There are a number of theoretical predictions for astrophysical and cosmological objects, which emit high frequency ($10^6-10^9$~Hz) Gravitation Waves (GW) or contribute somehow to the stochastic high frequency GW background. Here we propose a new sensitive detector in this frequency band, which is based on existing cryogenic ultra-high quality factor quartz Bulk Acoustic Wave cavity technology, coupled to near-quantum-limited  SQUID amplifiers at $20$~mK. We show that spectral strain sensitivities reaching $10^{-22}$ per $\sqrt{\text{Hz}}$ per mode is possible, which in principle can cover the frequency range with multiple ($>100$) modes with quality factors varying between $10^6-10^{10}$ allowing wide bandwidth detection. Due to its compactness and well established manufacturing process, the system is easily scalable into arrays and distributed networks that can also impact the overall sensitivity and introduce coincidence analysis to ensure no false detections. 
\end{abstract}

\maketitle


\section*{Introduction}
 
Gravitational radiation was first predicted by Einstein\cite{gw1} as a consequence of his General Theory of Relativity. Gravitational waves (GW) are the propagation of a wave of space-time curvature, and are generated by perturbations in massive systems. The lowest multipole of this type of radiation is the quadrupole. Even though astrophysical events are expected to emit massive energy fluxes in the form of gravitational radiation, they are yet to be directly detected. This is because gravity waves interact very weakly with matter. However, for many decades experimentalists have been pushing the limits of technology. Currently the free-mass laser interferometer detectors have been improved to a point, where they are expected to directly detect gravitational waves in the 0.1 to 1 kHz frequency band through the development of advanced LIGO\cite{advLIGO}.

The first gravitational wave detectors were based on the "Weber Bar", and required the monitoring of a high-$Q$ massive resonant system (resonant-mass detector). Such a system will change its state of vibration due to an incident gravitational wave of matched frequency and rely on ultra-sensitive transducers to readout the vibration. These transducers detect the displacement change of the system and convert it to an electronic signal. Since the first detector was built\cite{Weber:1960aa} technology improved rapidly over the years, with major projects in Italy, USA and Australia\cite{PhysRevLett.85.5046}. These detectors necessarily operated at low temperatures, and were successfully built at high sensitivity operating at 5 K down to 100 mK. The original transducers that Weber used were based on piezoelectricity, and later gap modulated displacement sensors were developed based on SQUID readouts\cite{Pleikies:2007aa} and low noise parametric systems\cite{PhysRevLett.74.1908}. These devices were optimised to detect millisecond bursts, typically produced by Supernovas with strain sensitivities of order $h_{1ms} >10^{-18}$ (or signal strain Fourier component $H>10^{-21}$ strain/Hz), but have generally been superseded by the laser interferometer detectors\cite{Whitcomb,Pitkin,Danilishin}. 

In this work we aim to revive the resonant-mass detector for the first cosmic search of high frequency gravitational wave radiation based on piezoelectric quartz Bulk Acoustic Wave (BAW) resonators. Despite dominance of the low frequency GW detection, this technology opens the way to test for known and unknown high frequency sources\cite{Cruise,Rudenko}. In general high frequency gravitational waves are thought to exist  over a broad range of frequencies up to $10^{10}$~Hz, but at a reduced amplitude compared to low frequency sources sensitive to LIGO. Such experiments could be interesting from two points of view: first, the high frequency region has physically understood processes of generation of GWs, second, such experiments can be regarded as tests for many emerging theories predicting GW radiation at such frequencies. The former mostly includes phenomena associated with discrete sources such as thermal gravitational radiation from stars\cite{Rudenko}, radiation from low mass primordial black holes\cite{Greene:2012aa,PBH,PhysRevD.58.063003}, gravitational modes of plasma flows\cite{PhysRevD.68.044017}, while the latter group is built up by cosmological sources including stochastic sources in the early universe\cite{Grish}, GW background from quintessential inflation\cite{PhysRevD.59.063505,PhysRevD.60.123511}, cosmic strings\cite{PhysRevD.54.7146,Brustein:1995aa}, dilation\cite{PhysRevD.47.1519}, pre-Big Bang scenarios\cite{Gasperini:2003aa}, superinflation in loop quantum gravity\cite{PhysRevD.79.023508}, post inflationary phase transitions\cite{PhysRevD.79.083519}, parametric resonance at the end of inflation or preheating\cite{PhysRevLett.99.221301,PhysRevLett.98.061302,PhysRevD.76.123517} and other predicted objects like brane-world black holes associated with extra dimensions\cite{PhysRevLett.94.121302,Clarkson} or clouds of axions\cite{PhysRevD.81.123530}. At least one of the hypothetical sources (due to the galactic centre shadow brane) comes within a factor 5 of the sensitivity of the single detector proposed in this work \cite{Cruise}.  Moreover, the sensitivity of the detector could be further improved via a variety of techniques to bridge this factor of 5. For example, the detector proposed in this work can be operated in the quantum limit, thus standard techniques to beat the quantum limit for the detection of a classical force can be used. Other ways to increase the sensitivity would be to use a larger resonant-mass structure, or an array of detectors. Thus, this technique will provide a valuable upper bound on such GW sources in the MHz frequency band and also provide an avenue for possible detection.

The technological advancement which allows this possibility is due to recent work on quartz bulk acoustic wave (BAW) resonators, which have been cooled to below 20 mK with outstanding acoustic properties\cite{galliou:091911,Goryachev1,ScRep,quartzPRL}. Also, they have proven to be compatible with SQUID amplifiers and offer quantum limited amplification at mK temperatures\cite{ourSQUID,Clarke}. The modes in these devices are naturally sensitive to gravitational waves at very high frequencies between 1 MHz to nearly 1 GHz with $Q$-factors of order $10^9$ and approaching $10^{10}$. The piezoelectricity and high acoustic $Q$-factor ensures significant "self" transductance of the BAW resonator with high electromechanical sensitivity without the necessity of an externally added transducer. These devices were originally developed for high stability frequency applications\cite{Salzenstein:2010aa,patrice2,Goryachev:2013ly} and more recently adapted at low temperatures for quantum information applications\cite{Goryachev1,quartzPRL,Goryachev:2014aa} and fundamental physics tests\cite{ScRep}. Indeed it seems like the quantum limited read out of these devices in the quantum ground state will soon be achieved. Temperatures are accessible where such modes will be in their ground state without the necessity of sideband cooling and near quantum limited SQUID and parametric amplifiers also exist at these frequencies\cite{SQUIDbook,Asztalos:2010aa, Yurke:1989aa,Clarke}. Here, we show that a sensitive gravitational wave detector of $10^{-21}$ strain per $\sqrt{\text{Hz}}$ can be realised with the present day technology in the frequency range of 1 to 1000 MHz. Currently there is no other technology capable of measuring such high frequency gravitational waves with such sensitivity, and a range of possible high frequency sources could exist. Only recently an idea of detecting medium frequency GW ($50-300$ kHz) using levitating optical resonators has been proposed\cite{PhysRevLett.110.071105}. In addition to that searches for GW background at $100$MHz have been considered with an interferometric detector\cite{PhysRevD.77.022002}. The system proposed in this work can potentially provide a larger frequency range and higher sensitivity for this purpose.

 \section{Acoustic Wave Distribution in a Curved Cavity}

\begin{figure}[htb!]
\centering
            \includegraphics[width=0.50\textwidth]{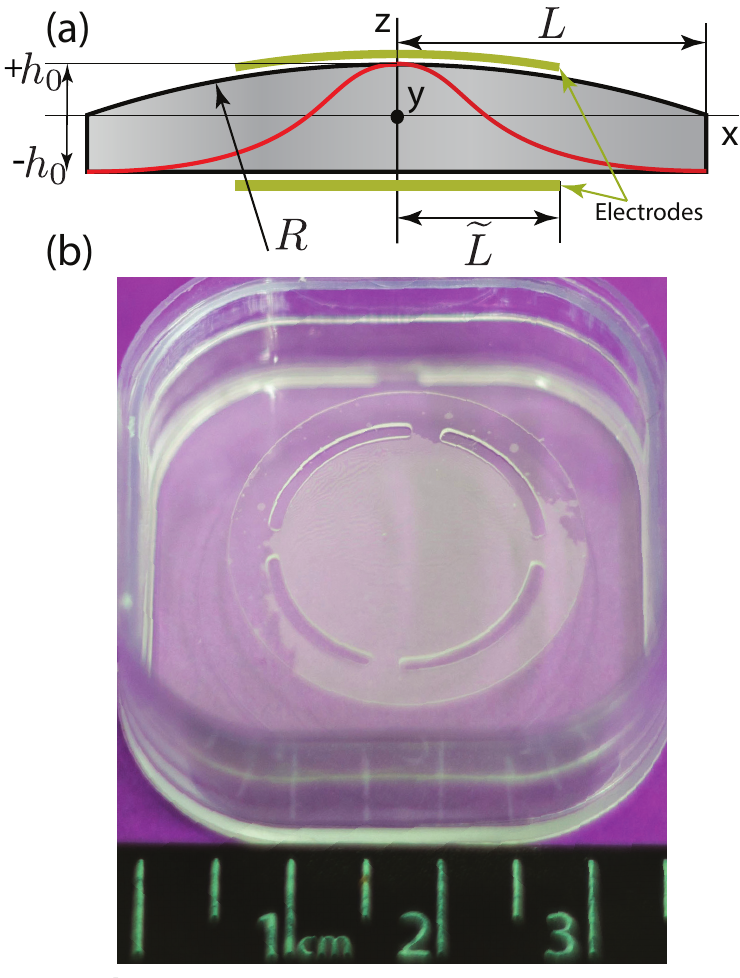}
    \caption{(a) Side view of a curved BAW plate cavity. Red curve shows typical distribution of mode displacement along the cut in the case $m=0$ and $p=0$. b) Top view photo of a BAW cavity plate in a plastic box (sample supplied by Serge Galliou)}%
   \label{F0041FS}
\end{figure}

The displacement distribution for the thickness modes of a plate resonator with curved surfaces can be calculated from the
Stevens-Tiersten theory\cite{Tiersten:1976hz,Tiers1, Shi:2014ac}. The theory establishes a partial differential equation for the dominant component of the displacement $u_d$, for the piezoelectric contoured BAW cavity with slowly varying thickness in the $x$-$y$ plane due to the large radius of curvature (Fig.~\ref{F0041FS})\cite{Tiers1}:
 \begin{equation}
\left. \begin{array}{ll}
\displaystyle \rho\ddot u_d +\frac{\pi^2 n^2 \hat{c}_{z}}{4h_0^2}\Big(1+\frac{x^2+y^2}{2Rh_0}\Big)u_d=\\
\displaystyle =M_n{\partial^2_{xx} u_d}+P_n\partial^2_{yy} u_d,
\end{array} \right. 
 \label{P117PP}
\end{equation}
where $n$ is the overtone number, $M_n$ and $P_n$ are parameters, which depending on material constants,
$R$ is the resonator plate radius of curvature, $2h_0\ll R$ is the resonator thickness, $\rho$ is the material mass density, $c_{z}$ is the effective elastic coefficients for the longitudinal mode. The acoustical cavity is also characterised by its length $L$ and electrode length $\widetilde{L}$ (Fig.~\ref{F0041FS}). 

The dominant component of the displacement $u_d$ is either along $x$, $y$ or $z$ axes (correspondingly $u_x$, $u_y$ or $u_z$) depending on the type of the thickness mode: longitudinal (A-mode), fast shear (B-mode) or slow shear (C-mode)\cite{eer}. Due to the higher sound velocity, the latter can be excited to much higher OTs and exhibits extremely high $Q$-factors\cite{landaurumer1}.

 Implying harmonic motion ${u}(x,y,z,t) = \overline{U}(x,y,z)e^{i\omega_{Xnmp}t}$, the eigensolutions of the homogenous problem corresponding to eq.~(\ref{P117PP}) can be estimated by
 \begin{equation}
 \left. \begin{array}{ll}
\displaystyle {U}_{Xnmp} =\sin\frac{n\pi z}{2h_0} e^{-\alpha n\pi\frac{x^2}{2}}H_m\big(\sqrt{\alpha n\pi}x\big),\times\\
 \displaystyle \times e^{-\beta n\pi\frac{y^2}{2}}H_p\big(\sqrt{\beta n\pi}y\big),
\end{array} \right. 
 \label{P118PP}
\end{equation}
where $X$ stands for a type of vibration (A, B or C modes), $m$ and $p$ are integer numbers characterising the acoustic wave distribution in the $x-y$ plane, $H_x$ is a Hermit polynomial and
 \begin{equation}
 \alpha^2 = \frac{\hat{c}_{z}}{8Rh_0^3M_p},\hspace{10pt} \beta^2 = \frac{\hat{c}_{z}}{8Rh_0^3P_n}.
 \label{P119PdP}
\end{equation}
It should be noted that typically, only resonances with the odd overtone number $n$ and even $m$ and $p$ numbers are piezoelectrically detectable. 

The angular frequencies of thickness modes of a curved plate is approximated as follows\cite{Goryachev:2014aa}:
 \begin{equation}
\omega_{nmp}^2 \approx \frac{n^2\pi^2\hat{c}_{z}}{4h_0^2\rho}\Big[1+\frac{\chi_x}{n}(2m+1)+\frac{\chi_y}{n}(2p+1)\Big]
 \label{P119PP}
\end{equation}
where $\rho$ is the resonator material density, $\hat{c}_{z}$ is a modified effective elastic constant for the given type of vibration, $\chi_x = \frac{1}{\pi}\sqrt\frac{2h_0M}{L\hat{c}_{z}}$ and $\chi_y = \frac{1}{\pi}\sqrt\frac{2h_0P}{L\hat{c}_{z}}$. For high-$Q$ BAW cavities the expression can be approximated by just the multiplier term before the square brackets, because in the limit of large $n$, $R\gg2h_0$  and low $m$ and $p$ numbers (usually both are zero) the last two terms in the expression are much less than 1.

For the case of the main modes ($m=0$, $p=0$), the effective mass is given by the expression\cite{Goryachev:2014aa}:
\begin{equation}
m_{n,0,0}  =\rho {\pi}{h_0L^2}\frac{\mbox{Erf}(\sqrt{2 n } \eta_x)\mbox{Erf}(\sqrt{ 2n} \eta_y)}{2\eta_x\eta_y n},
 \label{P016bDD}
\end{equation}
where $\eta_x = L \sqrt{\frac{\pi\alpha}{2}}$ and $\eta_y = L \sqrt{\frac{\pi\beta}{2}}$ are unitless trapping parameters.


For a given acoustic device the trapping parameters $\eta_x$ and $\eta_y$ could not be measured directly. Although it is possible to estimate these parameters based on measurements of spurious resonances of a certain overtone and resonator dimensions. 
Experimentally, it is often possible to determine resonance frequencies of the main OT $\omega_{n,0,0}$ and a few of its spurious resonance in particular $\omega_{n,2,0}$, $\omega_{n,0,2}$ and $\omega_{n,2,2}$. Utilising this information, two parameters of the general expression for the frequency (\ref{P119PP}) are
 \begin{equation}
\chi_x=n\frac{\omega_{n,2,0}^2-\omega_{n,0,0}^2}{5\omega_{n,0,0}^2-\omega_{n,2,2}^2},\chi_y=n\frac{\omega_{n,0,2}^2-\omega_{n,0,0}^2}{5\omega_{n,0,0}^2-\omega_{n,2,2}^2}.
 \label{PB001YY}
\end{equation}
These parameters calculated from the experimental data can be used to determine the ratios $\frac{M}{\hat{c}_z}$ for the given plate parameters $h_0$, $R$ and $L$. Substituted into (\ref{P119PdP}), this information leads to:
 \begin{equation}
\alpha= \frac{\chi_x}{2\pi h_0\sqrt{RL}},\beta= \frac{\chi_y}{2\pi h_0\sqrt{RL}}, 
 \label{PB002YY}
\end{equation}
which results in
 \begin{equation}
\eta_x = \frac{L}{2}\sqrt{\frac{\chi_x}{h_0\sqrt{RL}}},  \eta_y = \frac{L}{2}\sqrt{\frac{\chi_y}{h_0\sqrt{RL}}}.
 \label{PB002YY}
\end{equation}

\section{Acoustic Cavity Sensitivity to Gravitational Waves}

\subsection{Antenna Response} 

The vibration of a normal mode $\lambda=Xnmp$ of a gravitational wave antenna based on an acoustic cavity could be decomposed into\cite{thorne}:
 \begin{equation}
{u}_\lambda(\mathbf{x},t) = B_\lambda(t){U}_\lambda(\mathbf{x}), \int_\mathcal{V}dv{\rho}U_\lambda U_{\lambda^\prime}=\delta_{\lambda\lambda^\prime}{m}_\lambda
 \label{P120PdP}
\end{equation}
where $\mathcal{V}$ is the resonator volume. The second equation represents the normalisation condition which is merely the definition of the mode mass. The mode eigenfunction~(\ref{P118PP}) satisfies this condition. 

\begin{figure}[htb!]
\centering
            \includegraphics[width=0.5\textwidth]{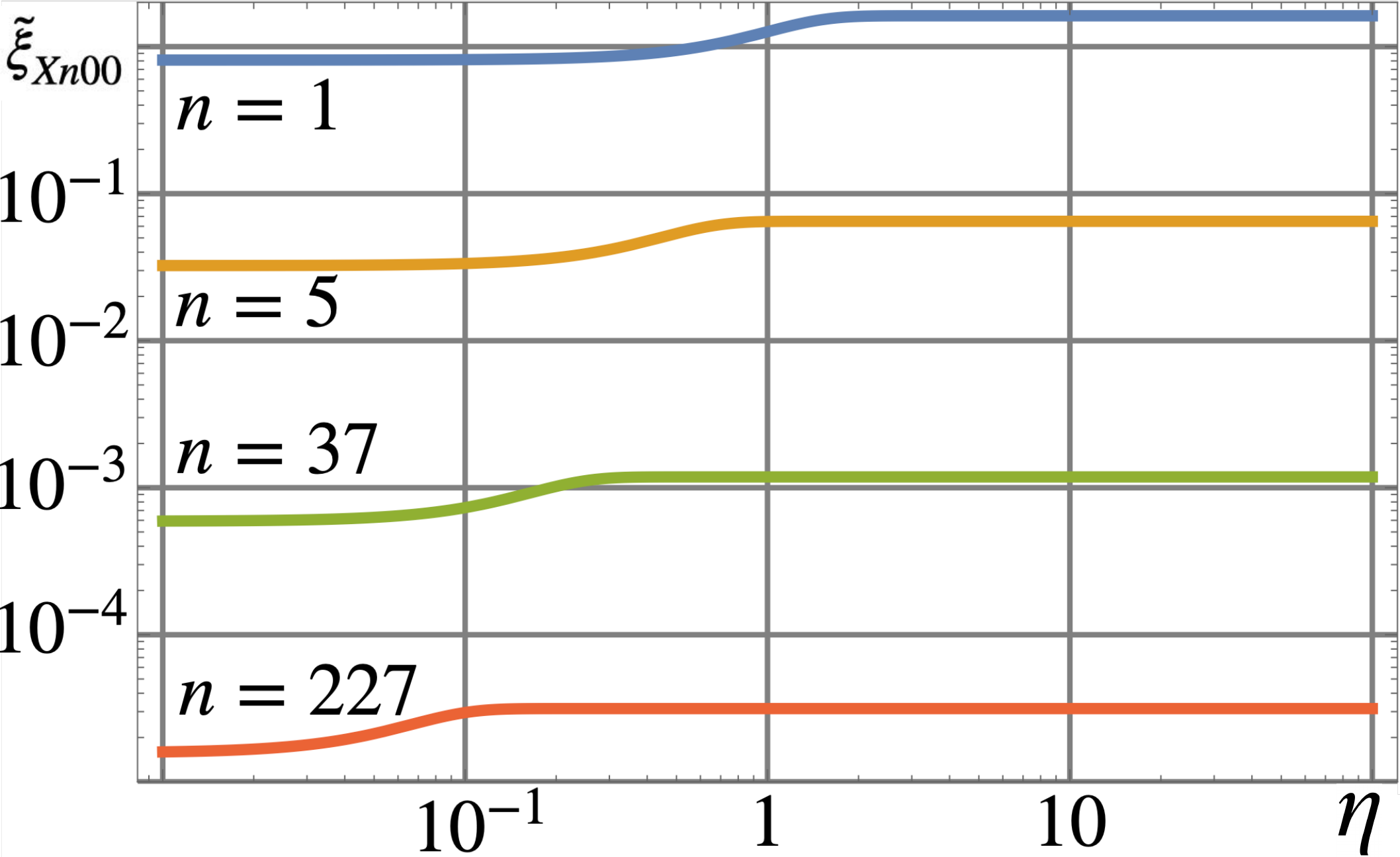}
    \caption{Gravitational wave sensitivity parameter $\widetilde{\xi}_{Xn00}$ as a function of the trapping parameter $\eta$.}%
   \label{F00111FS}
\end{figure}

According to \cite{Paik:1976aa,thorne}, the response $B(t)$ of the antenna
 to curvature tensor component $R^\alpha_{\beta\gamma\delta}$ is  
 \begin{equation}
 \left. \begin{array}{ll}
\displaystyle \ddot{B}_\lambda+\tau^{-1}_\lambda\dot{B}_\lambda+\omega_\lambda^2 B 
\displaystyle = -c^2R_{i0j0}\int_\mathcal{V}dv \frac{\rho}{m_\lambda}U_\lambda^i(\mathbf{x})x^j,
\end{array} \right. 
 \label{P121PP}
\end{equation}
where $\tau^{-1}_\lambda$ is the mode bandwidth and $x_j\in\{x,y,z\}$. The right-hand side provides the gravitational wave-cavity coupling:
 \begin{equation}
 \left. \begin{array}{ll}
\displaystyle \xi_\lambda ={h_0}\widetilde\xi_\lambda = \int_\mathcal{V}dv \frac{\rho}{m_\lambda} U_\lambda^i(\mathbf{x})x^j,
\end{array} \right. 
 \label{P122PP}
\end{equation}
that is nonzero only for $x_j = z$ for any $n$, $m$ and $p$. For the odd $n$th overtone with zero in-plane wave numbers, the sensitivity coefficient is
 \begin{equation}
 \left. \begin{array}{ll}
\displaystyle \widetilde\xi_{Xn00}=\frac{\xi_{Xn00}}{h_0} = \frac{16}{n^2\pi^2}\frac{\mbox{Erf}(\sqrt{ n} \eta_x)\mbox{Erf}(\sqrt{n} \eta_y)}{\mbox{Erf}(\sqrt{ 2n} \eta_x)\mbox{Erf}(\sqrt{ 2n} \eta_y)},
\end{array} \right. 
 \label{P123PP}
\end{equation}
This result suggests that the only dimension of the cavity sensitivity depends on is its thickness $h_0$. Other dimensions enter the results only through the dimensionless trapping parameter $\eta$. Assuming $\eta_x \approx \eta_y = \eta$, the sensitivity of various overtones of a BAW cavity is shown in Fig.~\ref{F00111FS}. 

\begin{figure}[htb!]
\centering
            \includegraphics[width=0.5\textwidth]{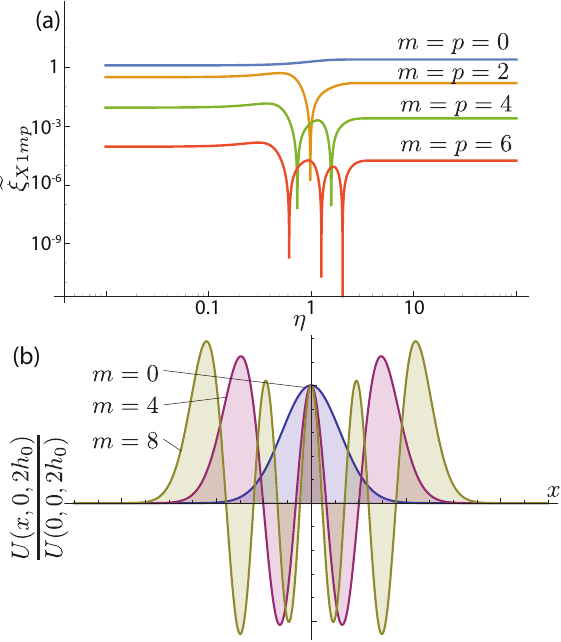}
    \caption{(a) Gravitational wave sensitivity $\widetilde{\xi}_{1mp}$ as a function of the trapping parameter $\eta$. The shaded area shows typical values for the well-trapped modes. (b) Normalised wave distribution along $x$.}%
   \label{F00112FS}
\end{figure}

The case of nonzero numbers $m$ and $p$ is shown in Fig.~\ref{F00112FS}, (a).  The same plot, subfigure (b), demonstrates the wave distribution along one of the plane coordinates. 

\subsection{Strain Sensitivity}

Near the quantum limit of operation, the single sided spectral density of the strain noise due to the Nyquist spectral density of force fluctuations acting on the antenna is given by\cite{Tobar:1995aa}:
 \begin{equation}
 \left. \begin{array}{ll}
\displaystyle \sqrt{S_h^+(f)} =\frac{2}{{\pi h_0}\widetilde\xi_\lambda f}\sqrt{\frac{E _\lambda}{m_\lambda Q_\lambda \omega_\lambda}} ~~ [\text{strain}/\sqrt{Hz}],
\end{array} \right. 
 \label{P334PP}
\end{equation}
where $E_\lambda$ is the Nyquist noise energy of the mode given by;
\begin{equation}
 \left. \begin{array}{ll}
\displaystyle E_\lambda = \chi_\lambda  k_BT_\lambda,
\end{array}  \right. 
 \label{P334PP2}
\end{equation}
where $T_\lambda$ is the mode temperature of overtone $\lambda$, $k_B$ is the Boltzmann constant and $\chi_\lambda$ is given by the Callen-Welton theorem\cite{IEEEnoise,PhysRev.83.34}:

\begin{equation}
 \left. \begin{array}{ll}
 \chi_\lambda = \hbar\omega_\lambda\beta \Big[\frac{1}{\exp(\hbar\omega_\lambda\beta)-1}+\frac{1}{2}\Big] 
\end{array}  \right. 
 \label{P334PP3}
\end{equation}
where $\beta_\lambda = 1/(kT_\lambda)$. Here the last term accounts for the vacuum fluctuations.

\begin{figure}[htb!]
\centering
            \includegraphics[width=0.5\textwidth]{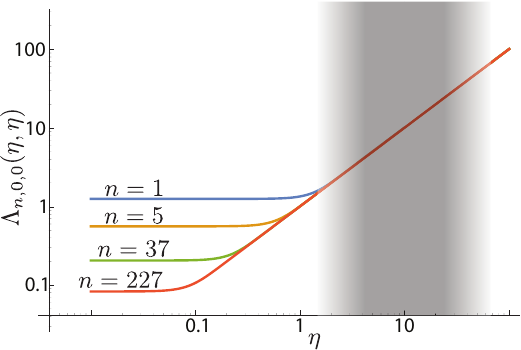}
    \caption{Geometry coefficient $\Lambda_{n,0,0}(\eta,\eta,1)$. Note that modes with $\eta<1$, quality factor is degraded by the clamping losses.}%
   \label{F00119FS}
\end{figure}

Assuming the ultimate limit of $Q$-factor, i.e. Landau-Ruomer dissipation, $Q_\lambda(\omega)=\text{const}$, the detection condition reduces to 
 \begin{equation}
 \left. \begin{array}{ll}
\displaystyle  \sqrt{S_h^+} =n\sqrt{\frac{kT\chi}{L^2 Q}\sqrt{\frac{\rho}{\hat{c}_z^3}}}\frac{\sqrt{\eta_x\eta_y\text{Erf}(\sqrt{2n}\eta_x)\text{Erf}(\sqrt{2n}\eta_y)}}{\text{Erf}(\sqrt{n}\eta_x)\text{Erf}(\sqrt{n}\eta_y)},
\end{array} \right. 
 \label{P335PP}
\end{equation}
As a numerical example, we consider a state-of-the-art acoustic cavity used to excite extremely high OTs as detailed in refernece\cite{quartzPRL}.
For this quartz device $L = 1.5\cdot 10^{-2}$~m, $\hat{c}_{z}\approx105$~GPa (could be varied by changing the cut), $\rho = 2643\frac{\mbox{kg}}{\mbox{m}^3}$, $h_0=5\cdot 10^{-4}$~m. The material parameter $\hat{c}_{z}$ is calculated to give the fundamental frequency of the quasi-longitudinal mode $f_{\mbox{fund}}=3.138$~MHz. The $Q$ factor could exceed $10^9$ at $T=20$mK. The resulting single sided power spectral density of the strain sensitivity is 
 \begin{equation}
 \left. \begin{array}{ll}
 \displaystyle  \sqrt{S_h^+} =n~4.3\cdot10^{-23}\frac{\sqrt{\eta_x\eta_y\text{Erf}(\sqrt{2n}\eta_x)\text{Erf}(\sqrt{2n}\eta_y)}}{\text{Erf}(\sqrt{n}\eta_x)\text{Erf}(\sqrt{n}\eta_y)}\\ \displaystyle = n~4.3\cdot10^{-23} \Lambda_{n,0,0}(\eta_x,\eta_y),\frac{\text{[strain]}}{\sqrt{\text{Hz}}}
\end{array} \right. 
 \label{P336PP}
\end{equation}
where the coefficient $\Lambda_{n,0,0}(\eta_x,\eta_y)$ is shown in Fig.~\ref{F00119FS}. The result suggests that for large enough trapping $\eta$, the geometric factor is independent of the overtone number and thus of frequency. Thus, it is possible to cover a large frequency range with modes sensitive to the gravitational waves. Note that for a resonator with no curvature, the trapping vanishes and $\Lambda_{n,0,0}(\eta_x,\eta_y)$ coefficient approaches unity.

\begin{figure}[htb!]
\centering
            \includegraphics[width=0.4\textwidth]{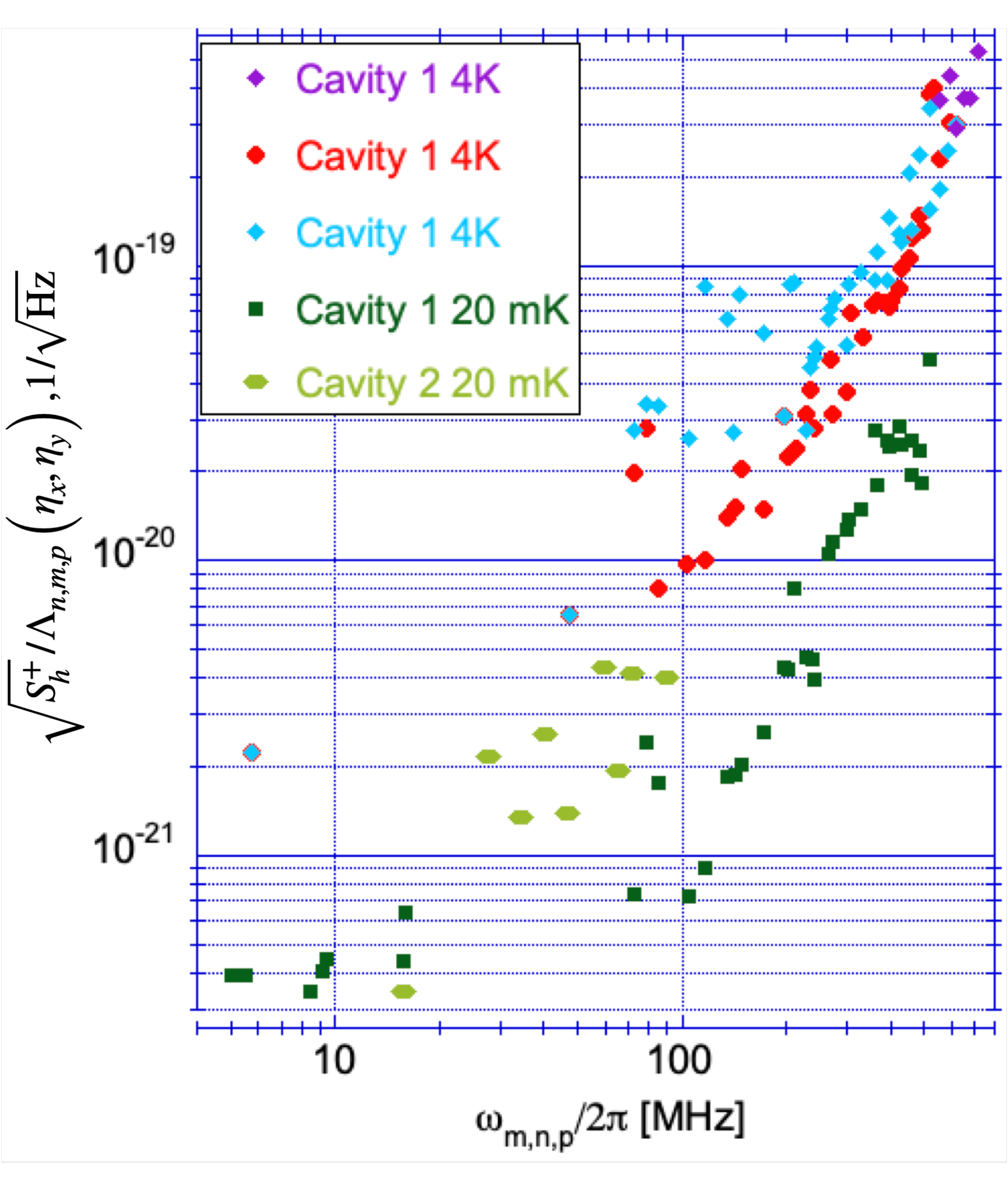}
    \caption{Corrected normalized single sided power spectral density of the strain sensitivity for various OTs of the longitudinal mode of two acoustical cavities at 4 K and 20 mK.}
   \label{A001AS}
\end{figure}

For actual cryogenic acoustic cavities, the $Q(\omega)=$const condition is not always fulfilled due to domination of other loss sources\cite{quartzPRL}. So, we estimate sensitivities for two devices that have been characterised at 4K and 20mK: Sample 1, $1.08$ mm thick, $13$ mm diameter electrode-separated disk cavities initially designed to sustain shear vibration of $5$~MHz at room temperature (manufactured by BVA Industrie)\cite{galliou:091911,Goryachev1,ScRep}; $1$ mm thick, $30$ mm diameter electrode-separated disk cavities with higher grade surface polishing initially designed to sustain shear vibration of $5$~MHz at room temperature (manufactured by Oscilloquartz SA)\cite{quartzPRL,ScRep}. The resulting comparison is shown in Fig.~\ref{A001AS}.

\subsection{System Calibration}

Acoustical vibration of a plate resonator $u$ is usually transformed to electrical current $I$ using piezoelectric properties of crystals. The direct calculation of the electromechanical coupling from the first principles then requires knowledge of the effective component of the crystal piezoelectric tensor $e_{\mbox{eff}}$:
\begin{equation}
I=-\int\limits_{\mathcal{A}_e}\partial_tD_zds = -e_{\mbox{eff}}\int\limits_{\mathcal{A}_e}\partial^2_{tz}u(x,y)ds
 \label{R120DD}
\end{equation}
where $\mathbf{D}$ is the displacement vector and $\mathcal{A}_e$ is electrode area.
Although, material parameters, for quartz in particular, are not known for cryogenic temperatures. Nevertheless, values of coupling could be derived based on impedance analysis of a BAW resonator. Such measurements provide three system parameters specific to each mode: resonance frequency $\omega_\lambda$, quality factor $Q_\lambda$ and motional resistance $R_\lambda$. This also can be done in an alternative representation using an electrical equivalent circuit with inductance $L_\lambda$, capacitance $C_\lambda$ and resistance $R_\lambda$. The electrical impedance of the mode $\lambda$ is then given as $Z_\lambda=j\omega L_\lambda+R_\lambda+\frac{1}{j\omega C_\lambda}$. Since only low overtone modes are considered in the present analysis, the influence of the shunt capacitance could be neglected. 

Equation (\ref{R120DD}) can be reduced to a simple charge form $q=\kappa_\lambda u$ assuming linearity of coupling. Here $\kappa_\lambda$ is the electromechanical coupling coefficient to be determined. Employing analogy between mechanical and electrical equivalent models for an acoustic resonance, it can be found that
\begin{equation}
M_\lambda = \kappa_\lambda^2 L_\lambda, k_\lambda = \frac{\kappa_\lambda^2}{C_\lambda},
 \label{R920DD}
\end{equation}
where $k_\lambda$ is an effective spring constant. It can be demonstrated that 
\begin{equation}
\kappa_\lambda^2=\frac{\omega_\lambda M_\lambda}{Q_\lambda R_\lambda}
 \label{R921DD}
\end{equation}
where the only parameter that cannot be measured directly with the impedance analysis is the mode mass $M_\lambda$ that can be estimated based on the acoustic wave distribution (\ref{P016bDD}). The parameter $\kappa_\lambda^2$ is typically small for quartz BAW devices\cite{Tiers1,Earness2}. It can be of the order of magnitude $\kappa_\lambda^2\sim10^{-4}-10^{-5}$~C$^2/$m$^2$, since typically $10$~MHz mode exhibits $Q\sim10^8$ and $R_\lambda \sim10$~Ohm with about $1-0.1$~g mass.

\section{SQUID-based Signal Detection}

Due to its piezoelectric nature, a BAW plate cavity can be directly coupled to an electronic circuit\cite{Goryachev:2014aa}. For cryogenic operation, the choice naturally falls on superconducting quantum interference devices due to their low noise. The successful observation of Nyqvist noise in cryogenic BAW cavity has been recently demonstrated\cite{ourSQUID}. This section discusses the noise limitations associated with this detection technique.

It has to be noted that the same BAW cavity can be probed by optomechanical methods \cite{Goryachev:2014aa} if a mirror coating is introduced. Optical probing of BAW quartz resonators has already been used for time-keeping and material study applications \cite{Dieulesaint:1982aa}. 

\begin{figure}[htb!]
\centering
            \includegraphics[width=0.5\textwidth]{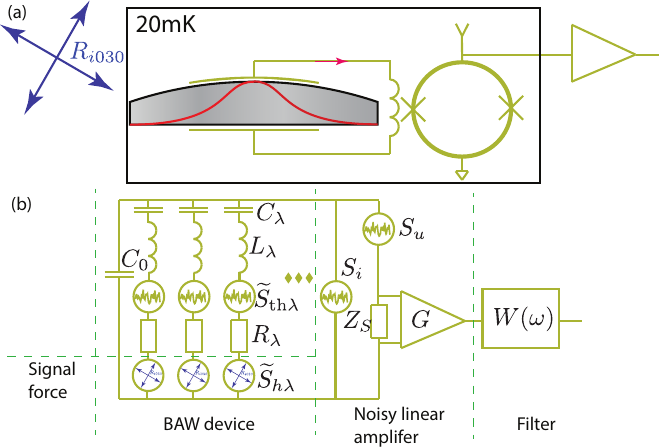}
    \caption{(a) Experimental setup. (b) Equivalent electrical model. }%
   \label{F00132FS}
\end{figure}

A typical gravitational-wave detection setup involving a resonant-mass antenna consists of three main parts: a resonant antenna, a noisy amplifier and a filter\cite{Giffard:1976aa,Price:1987aa}. The proposed experimental setup and the equivalent circuit model are shown in Fig.~\ref{F00132FS}. The first amplification stage of a detector is based on a near-quantum-limited SQUID amplifier that have already been employed for dark-matter search experiments\cite{Clarke}. 
To calculate the sensitivity of the whole setup, noise analysis of the dominant noise sources attributed to the Nyquist noise of the antenna and the amplifier must be considered. 

\subsection{SQUID Backaction Noise}

In this section we consider the amplifier Nyquist noise acting back on the antenna. The voltage noise acting back on the mode can be written as:
\begin{equation}
V_Q = \frac{Z_0Z_\lambda Z_S}{Z_SZ_0-Z_\lambda(Z_S+Z_0)}\big(I_s+\frac{V_s}{Z_S}\big) = Z_{BA}\big(I_s+\frac{V_s}{Z_S}\big),
 \label{FF01IO}
\end{equation}
where $Z_S$, $V_S$ and $I_S$ are input impedance, voltage and current noise of the SQUID amplifier and $Z_0$ is the impedance of the shunt capacitance $C_0$. The corresponding spectral density of the force fluctuations (measured in N$^2/$Hz) are
\begin{equation}
S_f =  \kappa_\lambda^2\big|Z_{BA}\big|^2S_{i}.
 \label{FF02IO}
\end{equation}
The back action impedance $Z_{BA}$ approaches $Z_\lambda$ in the limit $Z_S\rightarrow \infty$, $Z_0\rightarrow \infty$.

\begin{figure}[htb!]
\centering
            \includegraphics[width=0.5\textwidth]{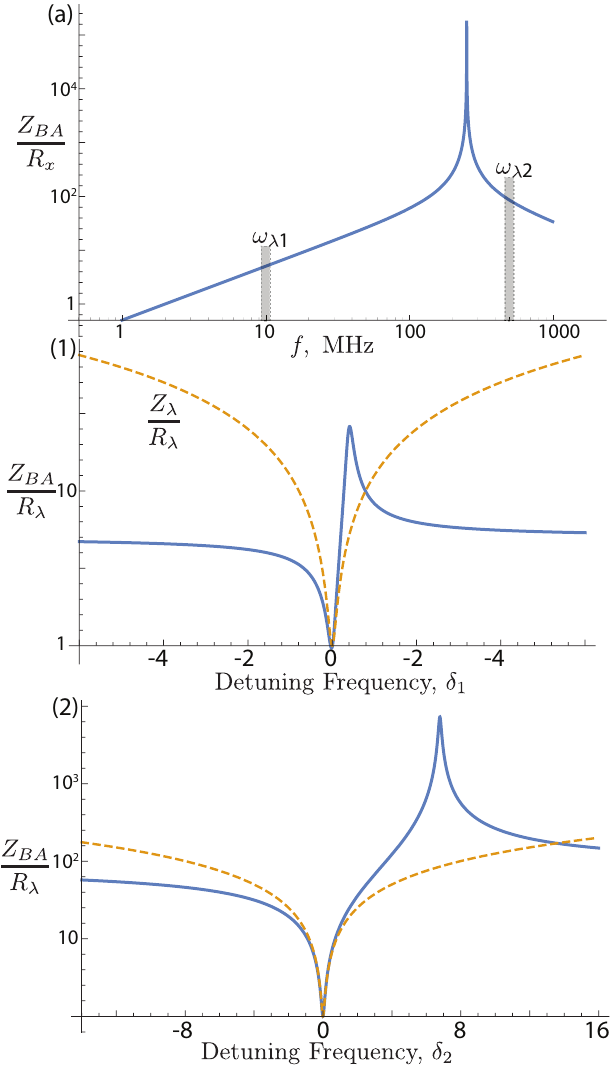}
    \caption{Backaction impedance $Z_{BA}$ (solid curves) in fractional units of $R_\lambda$. The detuning frequency given in number of bandwidths $\delta_i=\frac{\omega-\omega_{\lambda i}}{\omega_{\lambda i}}Q_\lambda$ is calculated for two resonance frequencies $\omega_{\lambda1}$ and $\omega_{\lambda2}$. Dashed curves demonstrate the normalised impedance of motional branches of he equivalent model. Note that the impedance drops to $R_\lambda$ at the mechanical resonances.}%
   \label{F00134FS}
\end{figure}

To estimate frequency dependence of $Z_{BA}$ impedance, we consider typical equivalent electrical parameters of an acoustic resonance ($R_\lambda=5$~Ohm, $L_x=1$~H, $C_0=1$~pF) and input coil of the SQUID amplifier ($L_S=400$~nH). From the calculations (see Fig.~\ref{F00134FS}) it is apparent that { the backaction impedance value exactly at the mechanical resonance equals to the motional impedance of an acoustic mode $R_\lambda$. This makes it independent of the resonance between the shunt capacitance $C_0$ and the input inductance of a SQUID amplifier as demonstrated in Fig.~\ref{F00134FS} (1) and (2).}  

{Result (\ref{FF02IO}) could be compared to the Nyquist force noise of the mechanical oscillator itself. The influence of these two types of noises is equal when
\begin{equation}
S_i =  4kT\chi\frac{R_\lambda}{|Z_{BA}|^2},
 \label{FF02IiO}
\end{equation}
which at mechanical resonance gives $S_i =  \frac{4kT\chi}{R_\lambda}$, i.e. the equivalent of noise corresponding to the mode equivalent resistance over the unit bandwidth. In other words, the SQUID backaction noise could be neglected if it is generated over the resistance less than that of the BAW mode.}


\subsection{SQUID Additive Noise}

Analysing the detector equivalent circuit (Fig.~\ref{F00132FS}, (b)), the spectral density of the signal at the system output is found as follows:
\begin{equation}
 \left. \begin{array}{ll}
 \displaystyle  S_{\mbox{out}} = W^2G^2\Big[\Big|\frac{Z_0Z_S}{Z_0Z_S-Z_\lambda(Z_0+Z_S)}\Big|^2 \widetilde{S}_\lambda+\\
\displaystyle+\Big|\frac{Z_0Z_\lambda}{Z_0Z_S-Z_\lambda(Z_0+Z_S)}\Big|^2 {S}_u\Big],
\end{array} \right. 
 \label{FF02CO}
\end{equation}
where $\widetilde{S}_\lambda$ is the PSD due to signal from the antenna and ${S}_u$ is due to the additive noise of the SQUID amplifier itself. From this relation the minimal detectable signal PSD (measured in m$^2/$Hz) is founds as
\begin{equation}
 \displaystyle  {S}_\lambda =\frac{1}{\kappa_\lambda^{2}{\omega_\lambda^2}{\big|Z_\lambda\big|^2} }\widetilde{S}_\lambda= \frac{{S}_u}{\kappa_\lambda^{2}{\omega_\lambda^2}{\big|Z_S\big|^2} }=\frac{{S}_\phi}{\kappa_\lambda^{2}{\big|Z_S\big|^2} },
 \label{FF04CO}
\end{equation}
where the flux noise of a squid amplifier $\sqrt{S_\phi}$ can be as low as $10^{-6}~\phi_0/\sqrt{\text{Hz}}$ (at $4$K) with $\phi_0=2.068\times10^{-15}$~Wb being the flux quantum\cite{ourSQUID}. { This gives flux noise $\sqrt{S_\phi}$ approaching $2.1\times10^{-21}$~Wb$^2/$Hz. With this noise parameter and SQUID input inductance $L_S=400~$nH and $\kappa_\lambda^2 = 10^{-4}$~C$^2/$m$^2$, the displacement sensitivity $\sqrt{S_\lambda}$ can be estimated as low as $3.3\times10^{-19}$~m$/\sqrt{\text{Hz}}$ at $1$~MHz and $10^{-20}$~m$/\sqrt{\text{Hz}}$ at $1$~GHz. Note that in order to minimise this parameter, the input inductance of the SQUID has to be maximised.}

\begin{figure}[htb!]
\centering
            \includegraphics[width=0.49\textwidth]{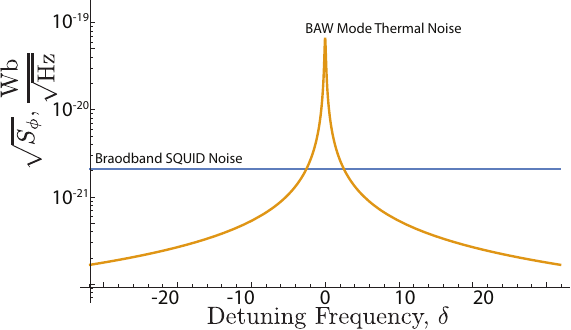}
    \caption{Comparison between SQUID additive (broadband) noise and BAW cavity mode Nyquist noise at the SQUID input around resonance frequency $\omega_\lambda$. The frequency scale is given in number of resonance bandwidths $\delta=\frac{\omega-\omega_{\lambda}}{\omega_{\lambda}}Q_\lambda$. }%
   \label{F00137FS}
\end{figure}

{ Similar to the case of the backaction noise, result (\ref{FF04CO}) could be compared to the intrinsic acoustic mode Nyquist noise. In this case, the flux noise corresponding to the antenna Nyquist noise is
\begin{equation}
S_\phi =  4kT\chi\frac{|Z_{S}|^2}{R_\lambda}\Big|\frac{\tau_\lambda^{-1}}{s^2+\tau_\lambda^{-1}s+\omega_\lambda^2}\Big|^2
 \label{FF02IrO}
\end{equation}
where $s$ is the Laplace variable. The last fraction of the right hand side represents the mechanical resonance transfer function. The result expression can be represented in the following form:
\begin{equation}
 \left. \begin{array}{ll}
\displaystyle S_\phi =  4kT\chi\frac{|L_{S}|^2}{R_\lambda}\Big|\frac{\tau_\lambda^{-1}\omega}{\tau_\lambda^{-1}j\omega+\omega_\lambda^2-\omega^2}\Big|^2=\\
\displaystyle=4.4\times10^{-39}\Big|\frac{\tau_\lambda^{-1}\omega}{\tau_\lambda^{-1}j\omega+\omega_\lambda^2-\omega^2}\Big|^2,\frac{\text{Wb}^2}{\text{Hz}}.
\end{array} \right. 
 \label{FF02ItO}
\end{equation}}
This result could be understood as BAW resonance Nyquist noise at the input of the SQUID amplifier measured in flux units. Fig.~\ref{F00137FS} compares this noise with the SQUID noise specified by manufacturer and discussed above. 

\section{Discussion and Future Perspectives}

In the previous sections, we have estimated the spectral sensitivity of a gravitational wave antenna based on a single BAW cavity. Remarkably the sensitivity in terms of spectral strain sensitivity can be better than the massive 700 Hz to 1 kHz GW detectors developed in the 1990's \cite{Tobar:1995aa} but operational at MHz frequencies. Thus, we have shown that they are suitable for sensitive detection of gravitational waves in the MHz frequency range. The advantages of this system include: 1) Extremely high $Q$-factors at cryogenic temperatures\cite{ScRep} from 4K down to mK temperatures, 2) well established high precision technology, 3) the existence of very large number of sensitive modes that can be used to probe distinct frequency bands\cite{quartzPRL}, 4) compactness, 5) ease of piezoelectrical coupling to SQUID and parametric amplifiers\cite{ourSQUID}, 6) also there is the possibility of organising arrays of detectors to improve sensitivity and allow coincidence analysis, and 7) the possibility of designing larger cavities for lower frequency ranges.

It has to be emphasized that the proposed system is a multi-mode detector. Due to large number of very high-$Q$ modes with a similar sensitivity in a wide frequency range, the bandwidth of system is not limited by a bandwidth of a single mode. Rather, the BAW cavity probes the GW radiation in different sample points across the frequency range. And due to large bandwidth of a SQUID amplifier, all the information from all modes, will be available at the system output. This fact may be important for analysis of stochastic GW background. 

While this work estimates the sensitivity for the currently available state-of-the-art BAW technology, there is room to optimise this technology for the purpose of increasing the sensitivity to gravitational waves. One possibility is to arrange several BAW cavities in an array. This arrangement alongside with cryogenic operation ($\sim 20$mK) becomes possible due to compactness of the devices ($<5$~cm$^{3}$ with a vacuum can). Although identical BAW devices manufactured according to the same technological process cooled to cryogenic temperatures have slightly different frequencies (up to tens of kHz) due to small imperfections, this may lead to finer frequency coverage of the incoming gravitational signal. Also, due to the fact that the quartz BAW technology is already well established, it would be possible to create a large network of such detector arrays in different laboratories all over the world. This international network of GW detectors may be used to exclude false detection via coincidence analysis\cite{PhysRevLett.85.5046}. 

Another possibility is to design a special upscaled BAW cavity for lower frequency range. It has been recently propose to use high quality BAW quartz resonators as mass standards\cite{Vig:2013aa}. Although manufacturing of such device is associated with considerable technological difficulties, the existence of it is very beneficial for standard keeping applications allowing not only more accurate mass standard but also signal transferring due to mass lock to frequency. This 1-kg standard will have a significantly lower frequency around $100$ kHz range, but augmented active mass a least $10^3$ times that is beneficial for the sensitivity. In the same time, quality factor of such device at cryogenic temperature cannot be reliably predicted, although it is generally observed that larger mass systems exhibit higher values of the $Q$-factor\cite{ScRep}. So, $Q$-factors in the range $10^6-10^9$ can be expected. Moreover, as the authors\cite{Vig:2013aa} suggest the mass standards in different laboratories can be connected in a synchronised network that replicates the idea of the large network of BAW cavity based GW detector arrays. 

One of the motivation for the development of the cryogenic BAW cavity technology is related to the field of engineered quantum systems. Indeed, the BAW systems have proven to have the highest $Q$-factor at the ground state among mechanical systems\cite{Kippen,quartzPRL}. In the same time, these devices are the largest objects (gram scale) that have been cooled to the ground state with the masses well above the Plank mass. So, BAW technology has potential ability to the quantum mechanics and the theory of general relativity under the same experimental framework\cite{aspelmayer2}. 

\section*{Acknowledgements}
The authors thank Serge Galliou for his useful discussions on BAW resonators and Maurice Van Putten for useful discussions on detecting gravitational radiation at high frequencies. This work was supported by the Australian Research Council Grant No. CE110001013 and FL0992016. The authors would like to thank Andraz Omahen, Yiwen Chu and William Campbell, who helped us uncover three errors, which have been corrected in this article on the 15/11/23. An errata in PRD will follow.

\hspace{10pt}

\section*{References}
%

\end{document}